\documentclass[aps,twocolumn,superscriptaddress,prl]{revtex4-1}

\usepackage{bm}
\usepackage{graphicx}
\usepackage{subfigure}
\graphicspath{{figures/}}

\def\be{\begin{equation}}
\def\ee{\end{equation}}
\def\e#1{\label{#1}\end{equation}}
\def\bea{\begin{eqnarray}}
\def\eea{\end{eqnarray}}
\def\ea#1{\label{#1}\end{eqnarray}}

\def\bem#1{\begin{mathletters}\label{#1}}
\def\eml{\end{mathletters}}

\def\4#1{{\boldsymbol{#1}}}
\def\8#1{{\widetilde{#1}}}
\def\bse{\begin{subequations}}
\def\ese{\end{subequations}}

\newcommand{\simgeq}{\; \raisebox{-0.4ex}{\tiny$\stackrel
{{\textstyle>}}{\sim}$}\;}

\begin{document}

\title{Solid-state electronic spin coherence time approaching one second}

\author{N. Bar-Gill}
\affiliation{Harvard-Smithsonian Center for Astrophysics, Cambridge, MA 02138, USA}
\affiliation{Department of Physics, Harvard University, Cambridge MA 02138, USA}

\author{L. M. Pham}
\affiliation{School of Engineering and Applied Sciences, Harvard University, Cambridge, MA 02138, USA}

\author{A. Jarmola}
\affiliation{Department of Physics, University of California, Berkeley, CA 94720-7300, USA}

\author{D. Budker}
\affiliation{Department of Physics, University of California, Berkeley, CA 94720-7300, USA}
\affiliation{Nuclear Science Division, Lawrence Berkeley National Laboratory, Berkeley, California 94720, USA}

\author{R. L. Walsworth}
\affiliation{Harvard-Smithsonian Center for Astrophysics, Cambridge, MA 02138, USA}
\affiliation{Department of Physics, Harvard University, Cambridge MA 02138, USA}

%
%
%
%
%

\begin{abstract}
Solid-state electronic spin systems such as nitrogen-vacancy (NV) color centers in diamond are promising for applications of quantum information, sensing, and metrology. However, a key challenge for such solid-state systems is to realize a spin coherence time that is much longer than the time for quantum spin manipulation protocols. 
Here we demonstrate an improvement of more than two orders of magnitude in the spin coherence time ($T_2$) of NV centers compared to previous measurements: $T_2 \approx 0.5$ s at 77 K, which enables $\sim 10^7$ coherent NV spin manipulations before decoherence. We employed dynamical decoupling pulse sequences to suppress NV spin decoherence due to magnetic noise, and found that $T_2$ is limited to approximately half of the longitudinal spin relaxation time ($T_1$) over a wide range of temperatures, which we attribute to phonon-induced decoherence. Our results apply to ensembles of NV spins and do not depend on the optimal choice of a specific NV, which could advance quantum sensing, enable squeezing and many-body entanglement in solid-state spin ensembles, and open a path to simulating a wide range of driven, interaction-dominated quantum many-body Hamiltonians.
\end{abstract}

\maketitle

In recent years, the electronic spin of the negatively charged nitrogen-vacancy (NV) color center in diamond has become a leading platform for applications ranging from quantum information processing \cite{laddquantum2010} to quantum sensing and metrology \cite{taylor2008,wrachtrup_efield}. Importantly, the NV spin-state can be optically initialized and detected on a timescale of $\sim 300$ ns\cite{childress2006} and coherently driven at up to GHz rates~\cite{awschalomGHz}. Single NV centers in isotopically-engineered high-purity diamond can possess long electronic spin coherence times on the order of a few ms at room temperature~\cite{wrachtrup_ultralong_2009}. By applying dynamical decoupling sequences, similar coherence times can be achieved for single NV centers \cite{hanson2010,cory,naydenov} and NV ensembles \cite{pham_dd} in diamond containing higher impurity concentration. 

Here we apply dynamical decoupling techniques to ensembles of NV centers over a range of temperatures (77 K - 300 K) in order to suppress both decoherence \cite{pham_dd} and phononic spin relaxation \cite{budkerT1}. 
We demonstrate an extension of the NV spin coherence time ($T_2$) to $T_2 \approx 0.5$ s at 77 K, which corresponds to an improvement of more than two orders of magnitude compared to previous measurements \cite{wrachtrup_ultralong_2009,pham_dd} and is on par with the longest coherence times achieved for electronic spins in any solid-state system \cite{lyon_longT2_Si} and for nuclear spins in room-temperature diamond \cite{lukin_nuclear_memory}.
Over a wide range of temperatures we also find that the NV spin $T_2$ is limited to approximately half of the longitudinal spin relaxation time ($T_1$), $T_2 \approx 0.5 T_1$, a finding that could be relevant to other solid-state spin defects (such as P donors in Si). The present result of NV $T_2$ approaching one second, in combination with single NV optical addressability and the practicality and scalability of diamond, advance NV centers to the forefront of candidates for quantum information, simulation, and sensing applications \cite{laddquantum2010,taylor2008}.

\begin{figure}
\begin{center}
\includegraphics[width=0.98 \linewidth]{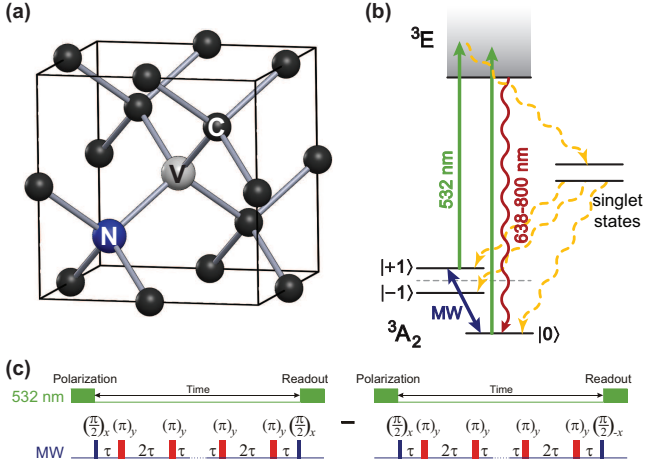}
\protect\caption{(a) Lattice structure of diamond with a nitrogen-vacancy (NV) color center. (b) Energy-level schematic of the negatively-charged NV center. (c) Measurement scheme: initialization and readout using 532 nm light, between which Carr-Purcell-Meiboom-Gill (CPMG) dynamical decoupling MW control sequences are applied. The sequence is repeated with the phase of the last $\pi/2$ pulse flipped (from $x$ to $-x$) for normalization (see text).} \label{fig:intro}
	\end{center}
\end{figure}

The NV center consists of a substitutional nitrogen atom and a vacancy occupying adjacent lattice sites in the diamond crystal [Fig. \ref{fig:intro}(a)]. The electronic ground state is a spin triplet [Fig. \ref{fig:intro}(b)], in which the $m_s=0$ and $\pm 1$ sublevels experience a $\sim 2.87$ GHz zero-field splitting. The NV center can be rendered an effective two-level system by applying a static magnetic field to further split the $m_s=\pm1$ states and addressing, e.g., the $m_s=0$ and $+1$ Zeeman sublevels. The NV spin can be initialized with optical excitation, detected via the state-dependent fluorescence intensity, and coherently manipulated using microwave pulse sequences \cite{childress2006}.

We performed measurements on an isotopically pure ($0.01\%$ $^{13}$C) diamond sample (Element Six) with nitrogen density $\sim 10^{15}$ $\rm{cm}^{-3}$ and NV density $\sim 3 \times 10^{12}$ $\rm{cm}^{-3}$. The sample was mounted inside a continuous-flow cryostat (Janis ST-500) with active temperature control (Lakeshore 331). Optical excitation was done with a 532 nm laser beam focused on the diamond surface through a microscope objective (NA = 0.6), and the resulting NV fluorescence was collected through the same objective and directed to a multimode fiber coupled to a single-photon counting module (Perkin-Elmer). The $\sim 30$ $\rm{\mu m}^3$ detection volume contained $\sim 100$ NV centers. Microwave control pulses were delivered to the NV spins using a $70$ $\rm{\mu m}$-diameter copper wire at the diamond surface.

We applied multi-pulse control sequences to decouple NV spins from the magnetic environment and thus extend the coherence time. Specifically, we used the Carr-Purcell-Meiboom-Gill (CPMG) pulse sequence \cite{cpmg} (Fig. \ref{fig:intro}(c)), with a varying number $n$ of $\pi$ control pulses. It has been shown previously \cite{pham_dd,naydenov,hanson2010,cory} that such pulse sequences are effective in extending the coherence time $T_2$ of NV spins. The coherence time increases as a power law of the number of pulses $T_2 \propto n^s$, where the specific scaling $s$ is determined mainly by the spin bath surrounding the NV \cite{bar-gill_spect_decomp} until spin-lattice relaxation begins limiting $T_2$ (assuming no pulse-error effects).

The coherence measurements were performed by applying the CPMG pulse sequence with the last $\pi/2$ pulse either along the x axis or the -x axis [Fig. \ref{fig:intro}(c)]. The two signals, labeled $r_1$ and $r_2$, were then subtracted and normalized to give the measurement results $m(t) = (r_1-r_2)/(r_1+r_2)$ such that common-mode noise is rejected. The final signal $m(t)$ is proportional to the coherence of the NV spin, defined as the magnitude of the off-diagonal density matrix element of the $m_s=0,+1$ two-level system $C(t) = Trace[\rho(t) S_x]$ [with $\rho(t)$ being the density matrix, and $S_x$ the transverse spin operator], at the end of the pulse sequence (before the final $\pi/2$ pulse). For a given pulse sequence (with $n$ pulses), the NV spin coherence as a function of time was measured by varying the free precession time between pulses and thus the total sequence time.

Measurements of the NV longitudinal spin relaxation followed a procedure similar to that used for the coherence measurements, with one of the two ``common-mode'' measurements initialized to $m_s=0$ (using an optical pumping pulse) and the other initialized to $m_s=+1$ (using an optical pumping pulse followed by a MW $\pi$ pulse). The normalized signal is proportional to the population difference between the spin states at time $t$, allowing measurement of the longitudinal spin relaxation and fitting to extract the $T_1$ relaxation time.

\begin{figure}
\includegraphics[width=0.98 \linewidth]{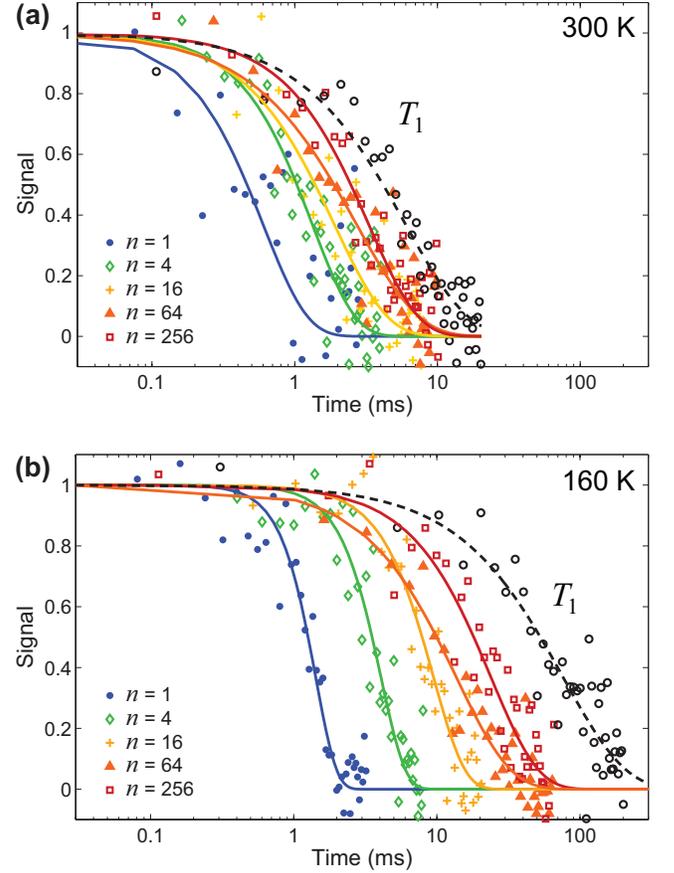}
\protect\caption{
Measured NV spin coherence decay used to determine $T_2$ for n-pulse CPMG sequences (colored data points and solid line fits to data), and longitudinal spin relaxation used to determine $T_1$ (black data points and dashed line fits to data), at (a) $T=300$ K and (b) $T=160$ K.}
\label{fig:res2temp}
\end{figure}

As an example of our data, we plot in Fig. \ref{fig:res2temp} the measured NV spin coherence as a function of time for Carr-Purcell-Meiboom-Gill (CPMG) pulse sequences\cite{cpmg} with different numbers of pulses $n$, at room temperature $T=300$ K (a) and at $T=160$ K (b). For each of these decoherence curves, we extracted $T_2$ by fitting the data to a stretched exponential function $f(t) = \exp{[-(t/T_2)^p]}$, where the $p$ parameter is related to the dynamics of the spin environment and inhomogeneous broadening due to ensemble averaging \cite{bar-gill_spect_decomp}. For both temperatures $T_2$ increases with the number of pulses in the CPMG sequence and is limited to about half of $T_1$: at 300 K, $T_2 = 3.3(4)$ ms and $T_1 = 6.0(4)$ ms; and at 160 K, $T_2 = 40(8)$ ms and $T_1 = 77(5)$ ms.

\begin{figure}
\includegraphics[width=0.98 \linewidth]{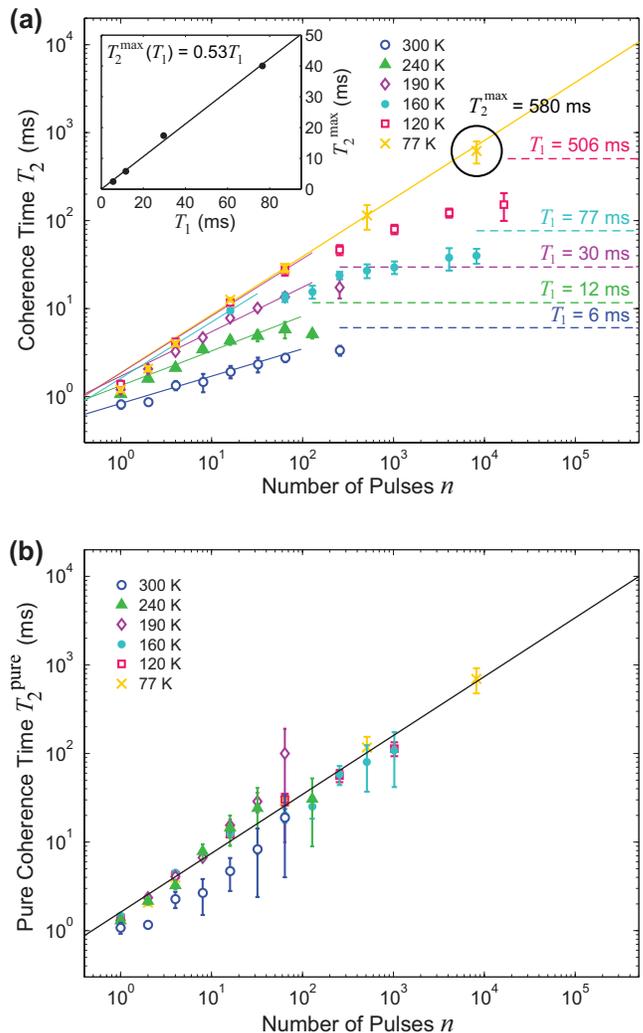}
\protect\caption{(a) Measured coherence time $T_2$ as a function of the number of pulses $n$ in the applied CPMG sequence, for temperatures from 77 - 300 K. For T=160 K and above, the corresponding measured longitudinal relaxation time $T_1$ is plotted as a dashed line. The black circle indicates the longest $T_2 = 580(210)$ ms for $n = 8192$ at 77 K. Inset: scaling of the maximum coherence time as a function of $T_1$: $T_2^{\mathrm{max}} = 0.53(2) T_1$. (b) NV coherence time with the phononic relaxation rate $[0.53(2) T_1]^{-1}$ subtracted to extract the ``pure'' spin-environment induced decoherence, with scaling behavior of $T_2^{\mathrm{pure}} \propto n^{0.67(3)}$ (see text).} \label{fig:allTemp}
\end{figure}

Figure \ref{fig:allTemp}(a) summarizes the measured dependence of $T_2$ on the number of CPMG pulses $n$ for several temperatures ranging from liquid nitrogen temperature (77 K) to room temperature (300 K). Notably, at liquid nitrogen temperature, an 8192-pulse CPMG sequence achieved a coherence time of $T_2 = 580$ ms, which is more than two orders of magnitude longer than previous NV $T_2$ measurements\cite{wrachtrup_ultralong_2009,pham_dd}.
Accumulated pulse errors limited sequences with larger $n$.
At $T=160$ K, a temperature that can be reached using thermoelectric cooling rather than cryogenic fluids, we found $T_2 = 40(8)$ ms, which is an order of magnitude longer than any previous NV $T_2$ measurement.
Thus, combined dynamical decoupling and thermoelectric cooling provides a practical way to greatly increase the NV spin coherence time, which could benefit many applications of NV ensembles, e.g. precision magnetometry \cite{pham_dd} and rotation sensing \cite{budker_gyro,paola_gyro}.

The $T_1$ relaxation time measured for each temperature is also indicated in Fig. \ref{fig:allTemp}(a) (except for 77 K, which had $T_1>10$ s). In the figure inset we plot the maximum $T_2$ achieved vs. $T_1$ at each temperature in the interval of 160 - 300 K and find $T_2^{\mathrm{max}} = 0.53(2) T_1$, which differs starkly from the previously expected $T_2$ limit of $2 T_1$ \cite{naydenov}. Under the assumption that spin-phonon coupling only causes spin-lattice ($T_1$) relaxation, one cannot recover the measured $T_2 \approx 0.5 T_1$ limit, even taking into account the possibility of unequal relaxation rates between the three ground-state NV spin sublevels (see Supplement). Therefore we conclude that spin-phonon coupling contributes significantly to NV spin decoherence \cite{hollenberg}. Such phonon-induced decoherence is generally relevant to any quantum system in which transitions between two levels can be driven by a two-phonon (Raman) process\cite{budkerT1}, and could play a role in the coherence properties of many other solid-state defects. At lower temperatures, to be studied in future work, spin-spin interactions should dominate both $T_1$ and $T_2$, and thus we expect $T_2^{\mathrm{max}}$ will deviate from the $\approx 0.5 T_1$ limit demonstrated here (see Supplement).

It is evident from Fig. \ref{fig:allTemp}(a) that the scaling of the coherence time with the number of CPMG pulses varies with temperature. In order to study the scaling behavior of the ``pure'' spin-environment-induced decoherence, we subtracted from the measured decoherence rate the temperature dependent phononic rate $1/[0.53(2) T_1]$ [see inset of Fig. \ref{fig:allTemp}(a)]. The corrected coherence time $T_2^{\mathrm{pure}}$ is plotted against the number of CPMG pulses $n$ in Fig. \ref{fig:allTemp}(b),
exhibiting striking universal behavior for all temperatures. We fit the corrected data to a power law scaling, and find $T_2^{\mathrm{pure}} \propto n^{0.67(3)}$. This value is consistent with the expected scaling power of $2/3$ for a Lorentzian noise spectrum of an electronic spin bath \cite{bar-gill_spect_decomp,sousa2009}. 

Even though our measurements were performed on an isotopically pure sample ($0.01\%$ $^{13}$C), we expect that similar results can be obtained for natural abundance diamond ($1.1 \%$ $^{13}$C), since the dynamical decoupling sequences we employ are also effective in suppressing dephasing caused by the nuclear spin bath\cite{cory,pham_dd}. However, special care must be taken in aligning the applied static magnetic field along the NV axis for natural abundance diamond since for ensembles of NVs in the presence of $^{13}$C nuclear spins, additional decoherence is caused by variations in the effective Larmor frequency of nearby nuclear spins due to magnetic field misalignment \cite{stanwix,mazePRB}.

The demonstrated improvement in coherence times for large ensembles of NV spins is directly relevant to enhanced metrology and magnetic field sensitivity, which scales as $(N_{\mathrm{NV}} T_2)^{-1/2}$, where $N_{\mathrm{NV}}$ is the number of sensing NVs \cite{taylor2008,pham_dd,budkerNVT2}. Thus, the hundred-fold increase in $T_2$ measured in this work at 77 K translates to a ten-fold improvement in magnetic field sensitivity. Using samples with higher densities of NVs could further improve this sensitivity, although at high NV concentrations ($\simgeq 1$ ppb) the coherence time may be limited by NV-NV interactions, since CPMG pulse sequences affect the NV spin and its NV spin-bath at the same time, thus canceling the decoupling effect. Other techniques, such as WAHUHA and MREV pulse sequences\cite{wahuha,mrev}, may be applied to address this issue.

We note that achieving long NV spin coherence times in diamond samples with high impurity concentration ($\sim 1$ ppm) is a crucial step toward creating non-classical states of NV ensembles. Such non-classical states could form the basis for high-sensitivity quantum metrology, potentially allowing significantly improved sensitivity and bandwidth \cite{oberthaler_nonlinear,oberthaler_homodyne}, and could also serve as a resource for quantum information protocols. To observe significant entanglement between neighboring NV centers, their decoherence rate must be small compared to the frequency associated with their interaction. For a realistic diamond sample with [N] $\sim 1$ ppm and [NV] $\sim 10$ ppb, coherence times larger than $\sim 50$ ms are needed for significant entanglement, which is within reach given the results presented here. For example, collective spin squeezing using $L_Z^2$ one-axis squeezing techniques\cite{ueda} could be created through application of pulse-sequences that average-out the $X,Y$ components of the spin-spin dipolar coupling\cite{cappellaro_squeezing}. Such pulse sequences can be straightforwardly applied in conjunction with the CPMG pulse sequences used here for extending the NV spin coherence time.

In conclusion, we demonstrated more than two orders-of-magnitude improvement in the coherence time ($T_2$) of ensembles of NV electronic spins in diamond compared to previous results: up to $T_2 \approx 0.5$ s, by combining dynamical decoupling control sequences with cryogenic cooling to 77 K; and $T_2 \simeq 40$ ms for temperatures achievable via thermoelectric cooling ($>$160 K). By studying the dependence of $T_2$ and of the NV spin relaxation time ($T_1$) on temperature, we identified an effective limit of $T_2 \approx 0.5 T_1$, which we attribute to phonon-induced decoherence. Given this limit, we expect that for low NV densities a $T_2$ of a few seconds should be achievable at liquid-nitrogen temperatures ($T = 77$ K).
 
The greatly extended NV spin coherence time presented in this work, which does not require an optimally chosen NV center, could form the building block for wide-ranging applications in quantum information, sensing, and metrology in the solid-state \cite{hanson_dd_gates,jero2008}. In particular, the fact that such long coherence times can be achieved with high-density ensembles of NVs suggests that spin squeezing and highly entangled states can be created, since $T_2>$NV-NV dipolar interaction time.
Finally, this work could provide a key step toward realizing interaction-dominated topological quantum phases in the solid-state, as well as a large family of driven many-body quantum Hamiltonians \cite{yao_topo}.


\paragraph{Acknowledgements}
This research has been supported by the DARPA QuASAR program, NSF, IMOD, and the NATO Science for Peace Program.
We gratefully acknowledge discussions with Ran Fischer, the provision of diamond samples by Element Six, and helpful technical discussions with Daniel Twitchen, Matthew Markham, Alastair Stacey, Keigo Arai, Chinmay Belthangady, David Glenn, and David Le Sage.

\bibliography{NV}

\end{document}